\documentclass[conference]{IEEEtran}
\IEEEoverridecommandlockouts
\usepackage{cite}
\usepackage{amsmath,amssymb,amsfonts}
\usepackage{algorithmic}
\usepackage{graphicx}
\usepackage{float}
\usepackage{textcomp}
\usepackage{xcolor}
\usepackage{subfigure}
\usepackage{url}
\usepackage{algorithm}
\usepackage{algorithmic}
\usepackage{makecell}

\usepackage{cite}
\usepackage{amsmath,amssymb,amsfonts}
\usepackage{algorithmic}
\usepackage{algorithm}
\usepackage{graphicx}
\usepackage{textcomp}
\usepackage{xcolor}
\usepackage{fancyhdr}
\def\BibTeX{{\rm B\kern-.05em{\sc i\kern-.025em b}\kern-.08em
    T\kern-.1667em\lower.7ex\hbox{E}\kern-.125emX}}
\begin{document}

\pagestyle{fancy}
\fancyhead[C]{This paper appears in IEEE International Conference on Communications, 2023.}
\title{Mobile Edge Adversarial Detection for Digital Twinning to the Metaverse with Deep Reinforcement Learning}

\author{\IEEEauthorblockN{Terence Jie Chua}
\IEEEauthorblockA{Graduate College\\Nanyang Technological University\\
terencej001@e.ntu.edu.sg }
\and

\IEEEauthorblockN{Wenhan Yu}
\IEEEauthorblockA{Graduate College\\Nanyang Technological University\\
wenhan002@e.ntu.edu.sg }
\and

\IEEEauthorblockN{Jun Zhao}
\IEEEauthorblockA{School of Computer Science \& Engineering\\ Nanyang Technological University\\
junzhao@ntu.edu.sg }
}

\maketitle
\thispagestyle{fancy}

\begin{abstract}
Real-time Digital Twinning of physical world scenes onto the Metaverse is necessary for a myriad of applications such as augmented-reality (AR) assisted driving. In AR assisted driving, physical environment scenes are first captured by Internet of Vehicles (IoVs) and are uploaded to the Metaverse. A central Metaverse Map Service Provider (MMSP) will aggregate information from all IoVs to develop a central Metaverse Map. Information from the Metaverse Map can then be downloaded into individual IoVs on demand and be delivered as AR scenes to the driver. However, the growing interest in developing AR assisted driving applications which relies on digital twinning invites adversaries. These adversaries may place physical adversarial patches on physical world objects such as cars, signboards, or on roads, seeking to contort the virtual world digital twin. Hence, there is a need to detect these physical world adversarial patches. Nevertheless, as real-time, accurate detection of adversarial patches is compute-intensive, these physical world scenes have to be offloaded to the Metaverse Map Base Stations (MMBS) for computation. Hence in our work, we considered an environment with moving Internet of Vehicles (IoV), uploading real-time physical world scenes to the MMBSs. We formulated a realistic joint variable  optimization problem where the MMSPs' objective is to maximize adversarial patch detection mean average precision (mAP), while minimizing the computed AR scene up-link transmission latency and IoVs' up-link transmission idle count, through optimizing the IoV-MMBS allocation and IoV up-link scene resolution selection. We proposed a Heterogeneous Action Proximal Policy Optimization (HAPPO)  (discrete-continuous) algorithm to tackle the proposed problem. Extensive experiments shows HAPPO outperforms baseline models when compared against key metrics.

\end{abstract}


\section{Introduction}
\textbf{Background. }Digital twinning is the keystone of the Metaverse~\cite{lee2021all}, in which real-world objects and events are mapped to and replicated in the virtual world. This opens doors to a myriad of possible applications which require real-time information of the physical environment, such as Augmented Reality (AR)-assisted driving. To facilitate AR-assisted driving capabilities, real-world scenes have to be uploaded to a central Metaverse Map Service provider (MMSP) which functions as a virtual reality host for geographical information. The physical world scenes have to be collated, aggregated to form a coherent database. Internet of Vehicles (IoVs) can then query information from the MMSP and this information can be displayed as AR scenes on the IoVs' windshield which provide drivers with comprehensive, real-time information such as directions and landmark information to assist their driving.

\textbf{Motivation. }The development of the new-age AR-assisted driving technology invites adversaries. Adversaries may physically paste adversarial patches on cars, signboards, traffic lights or on the roads with the intent to corrupt the physical world scene which is to be uploaded to the MMSP for the development of a centralize virtual map. A successful attack as such can have disastrous effects, in which the Metaverse Map scenes may reflect erroneous information which when queried by IoVs can result in misinformation and accidents.


\textbf{Compute intensive detection. }These adversarial patches are often inconspicuous~\cite{bai2021inconspicuous}, and a fairly high resolution image of the patch is required for patch detection. This makes real-time detection of adversarial patches compute intensive, and these adversarial patch detection task have to be offloaded to the Metaverse Map Sevice Provider Base Stations (MMBSs) for computation. However, the offloading of high-resolution physical world scenes may induce large uplink transmission latency, yet offloading low-resolution images substantially impairs the MMSPs' adversarial patch detection ability. Furthermore, too many IoVs allocated to an MMBS may result in sub-optimal performance and unreliability in the system. Hence, IoVs may be excluded from certain UL transmissions if the occasional exclusion of an IoV results in better system performance. The total number of exclusions (idle count) should be minimized to ensure that the MMSP obtains comprehensive and regular information update from the IoVs.

\textbf{Our Approach. }Hence, we proposed a Heterogenous Action Proximal Policy Optimization (HAPPO) algorithm to be employed within the MMSP orchestrator to tackle the optimization problem of (i) maximizing patch detection mean average precision (mAP) while minimizing the (ii) uplink latency and (iii) IoV idle count. The orchestrator consists of two agents, one to handle (1) discrete IoV-MMBS allocation and the other to handle (2) continuous physical scene resolution selection. Our HAPPO architecture follows the Centralized Training and Decentralized Execution (CTDE) framework~\cite{lowe2017multi}.

\subsection{Related work}


\textbf{Adversarial Patches. }
The detection of adversarial perturbations within images has been thoroughly studied~\cite{ma2018characterizing, zheng2018robust}. Many works~\cite{feinman2017detecting,gong2017adversarial,lee2018simple} in the field of adversarial detection have built classifiers to sift out  corrupted samples from natural (unperturbed or clean) samples. However, as physical attack’s practicality in real-world gains recognition, there is an increasing number of researches focused on developing better defenses against adversarial patch attacks. Recent works~\cite{arvinte2020detecting,xu2020lance}
 in adversarial defenses have been focused on adversarial detection. These works utilize heuristic-based approaches such as using wavelets~\cite{arvinte2020detecting} and Grad-CAM maps~\cite{xu2020lance} to differentiate between natural and adversarial samples.

\textbf{Metaverse applications. }Since the Metaverse is still relatively new, limited studies consider the IoT-Metaverse base station communication and computation framework. Chua~\textit{et~al.}~\cite{Chua2022} introduced a AR socialization over 6G wireless networks within the Metaverse scenario and proposed a deep RL approach to tackle it. Han~\textit{et~al.}~\cite{han2022dynamic} addressed resource allocation for the MEC of digital twinning of Internet of Things (IoT). Similarly, Ng~\textit{et~al.}~\cite{ng2022unified} tackled a resource allocation problem for the MEC of the Metaverse education sector using stochastic optimization.


\textbf{Resource Allocation. }Resource allocation for wireless networks have been thoroughly studied~\cite{liu2019resource2,ahsan2021resource,hieu2022joint}. Works such as those by~\cite{hieu2021optimal} utilized deep reinforcement learning approaches to allocate power for communications.

\textbf{Edge Computing. }Resource allocation and optimization of variables have been a long-standing concern in the field of mobile edge computing (MEC), and there have been several works~\cite{hu2018joint,liu2019resource} which presented edge computing problems and developed solutions to tackle their proposed problem. Some works have adopted deep reinforcement learning approaches to tackle optimization problems for mobile edge computing~\cite{huang2019deep,huang2018deep,huang2018distributed}.

\noindent\textbf{Contributions.} Our contributions are as follows:
\begin{itemize}
\item \emph{\textbf{Adversarial Detection in Defence of the digital twinning: }}
We present a novel mobile edge computing (MEC)-enabled adversarial patch detection for the defence of digital twinning to the Metaverse scenario, specifically in the context of AR-assisted driving.

\item \emph{\textbf{HAPPO approach to Asymmetric Joint Optimization Problem: }}We propose a Heterogeneous Action PPO, dual-agent (discrete-continuous) deep reinforcement learning-based IoV to MMBS orchestrator which aims to maximize adversarial patch detection mAP while minimizing total physical scene up-link transmission delay and IoV idle count.

\item \emph{\textbf{Superiority of HAPPO: }} We conducted experiments to compare the performance of our proposed HAPPO against other base-line algorithms and results demonstrate the effectiveness and superiority of our proposed method.

\end{itemize}

\begin{figure}[t]
    \centering
    \includegraphics[width=1\linewidth]{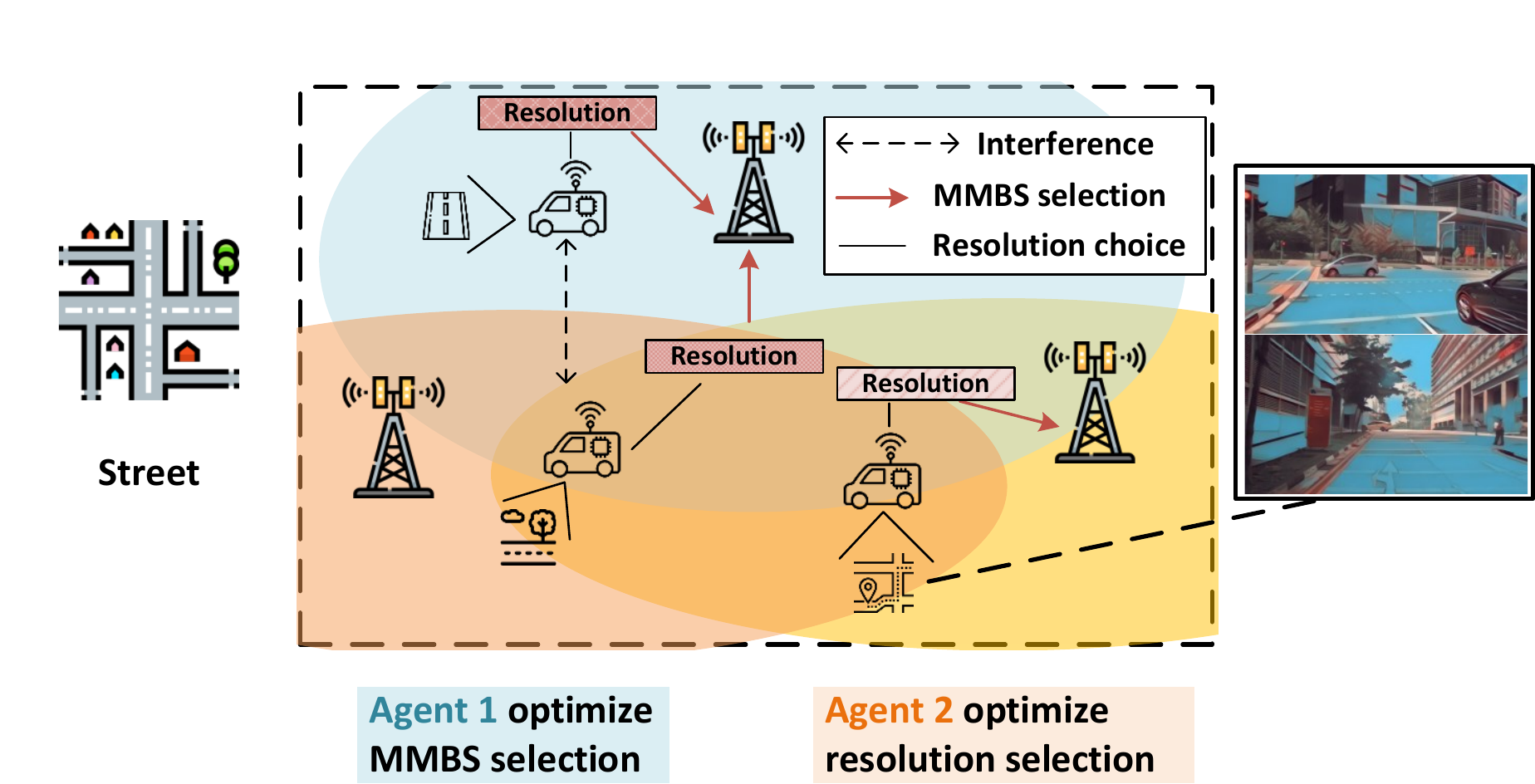}    \caption{System Model involving 2 agents to facilitate the adversarial detection offloading.}
    \label{fig:systemmodel}
    \vspace{-0.5cm}
\end{figure}

\section{Adversarial Patch}
\textbf{Adversarial Patch Attack. }\label{sub:patch_attack}
For simplicity, we inserted an adversarial patch, digitally, onto images of cars and roads, 2000 from each of the \textit{Stanford-Cars}~\cite{krause2013collecting} and \textit{nuImages}~\cite{nuscenes2019} datasets, to mimic the placement of physical adversarial patches onto the physical environment. We assigned square-shaped patches of size smaller than 2\% of the total image area randomly on our training dataset. We adopted the Projected Gradient Descent (PGD) patch attack~\cite{madry2017towards} as an example attack to be applied on our training set. The mechanism of PGD can be described as such: The algorithm aims to find a perturbation value to be added to the natural example, which maximizes the loss function value, under the constraints in which the norm of the perturbation value falls within a pre-defined threshold (shown in equation \ref{pgd_function}).
\begin{align}
    \max_{\lVert\zeta\rVert_\infty \leq \varphi} l(\mathcal{F}(\chi_0 + \zeta; \varpi), \Upsilon_0)
    \label{pgd_function}
\end{align}
\noindent where $\chi_0$ represents the natural example, $\Upsilon_0$ represents the original label, $\zeta$ is the perturbation value to be added, $\varpi$ is the model weights, $l$ is the loss function, and $\varphi$ represents the perturbation threshold value. $\mathcal{F}$ is the predictive function that maps the input to a prediction. Implementing it by iterative gradient descent, we have:
\vspace*{-0.5cm}

\begin{align}
    \chi_{t+1} = \chi_{t} + \kappa \cdot \text{sign}(\nabla_\chi l(f(\chi_t;\varpi),\Upsilon_0)) \label{iterative}
\end{align}

\noindent where $\chi_t$ refers to the current image state, and $\kappa$ is a scalar multiplier.

\begin{figure*}[t]
    \centering
    \includegraphics[width=0.8\linewidth]{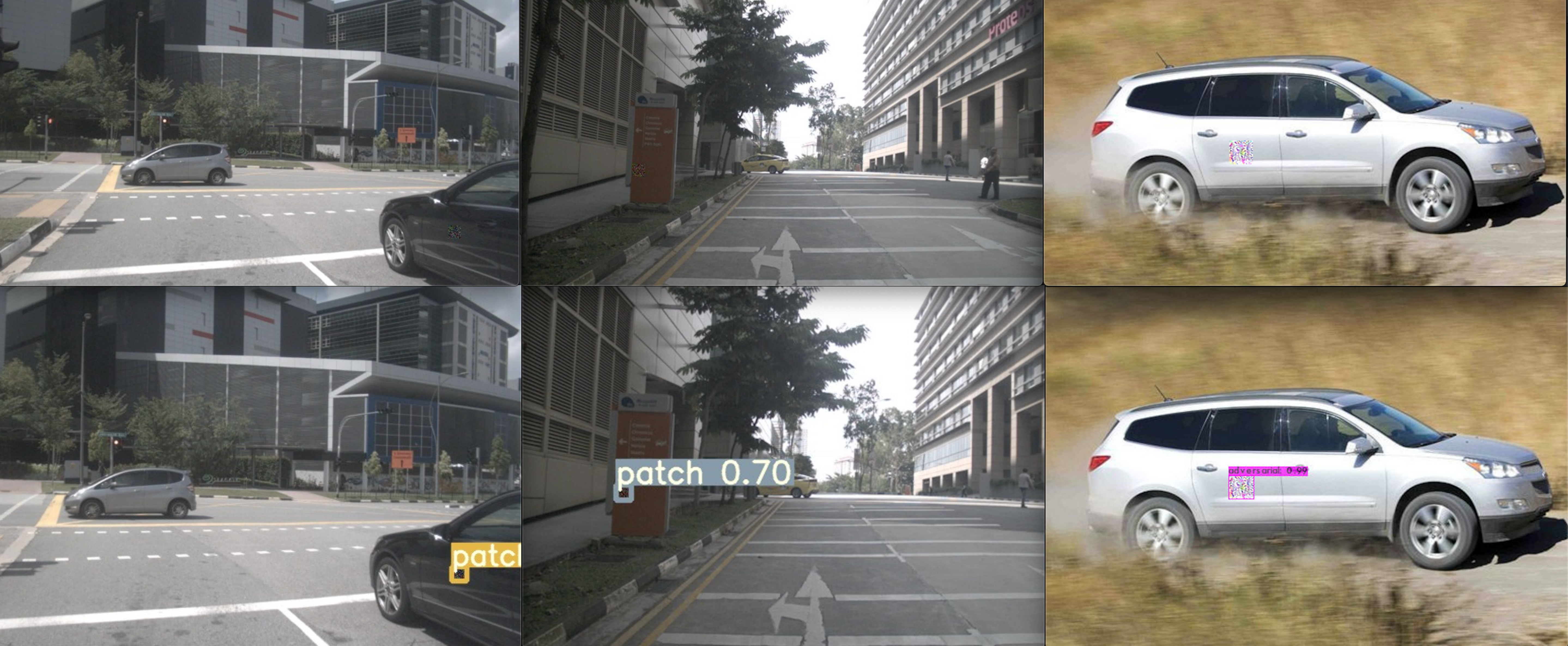}
    \caption{Detection of Adversarial Patches from \textit{nuImages} (left and middle)~\cite{nuscenes2019} and \textit{Stanford-Cars} (right)~\cite{krause2013collecting} datasets.}
    \label{fig:patches}
    \vspace{-0.5cm}
\end{figure*}

\textbf{Adversarial Patch Detection. }
\label{sub:patch_detect}
After the physical scenes with adversarial patches are offloaded from the IoVs to the MMBSs, the trained adversarial patch detectors on the MMBS will detect for adversarial patches on the uploaded physical world scenes. We adopted a cutting-edge object detector, pre-trained YOLOv4 with a CSPDarknet53 backbone~\cite{bochkovskiy2020yolov4} to detect the adversarial patches (shown in Fig.~\ref{fig:patches}). In order to reduce latency, lower resolution images may be transmitted to the MMBS, which consequently result in poorer patch detection mean average precision score (mAP). Vice versa, transmission of higher resolution images results in higher latency but better patch detection mean average precision score (mAP). mAP is a common performance metric used in object detection tasks, which takes the mean of average precision (AP) scores across different intersection over union (IoU) bounding box thresholds. In our work, we adopt the IoU threshold values from 0.5 to 0.95 in incremental steps of 0.05.

\section{System model}
\label{models}
In our system, $N$ AR vehicles from a set of $\mathcal{N}=\{1,2,...,N\}$ AR vehicles are capturing and uploading physical world scenes in real-time on the go, to Metaverse Map Service Provider Base Station (MMBSs) $\mathcal{M} = \{1,2,...,M\}$.  Each AR vehicle $i \in \mathcal{N}$ moves around a defined geographical space at random and uploads physical environment scenes to an MMBS (shown in Fig.~\ref{fig:systemmodel}). As high-resolution, large data-size physical world scenes are required for patch detection, there may be a hand-over of the physical scenes uploaded, from one MMBS to another, AR vehicles move within a defined space. Several MMBSs and AR vehicles are distributed geographically. These AR vehicles transmit the physical world scenes to the MMBS and may generate interferences that disrupts the effective signal between other AR vehicles and their assigned MMBS. In our paper, we consider intra-cell interference. Intra-cell interference in this context, refers to the signal interference caused by the transmissions of other AR vehicles on the same bandwidth, and are assigned to the same MMBS, as the AR vehicle of interest.

\textbf{Uplink Communication model. }\label{sub:Communication-Model}
Each AR vehicle from a set of $\mathcal{N}=\{1,2,...,N\}$ will be assigned an MMBS's downlink channel from a set of $\mathcal{M} = \{1,2,...,M\}$ MMBS.
The physical world scenes to be uploaded from the AR vehicles to the MMBS are of size $d^t = \{d^t_{1}, d^t_{2}, ..., d^t_{N}\}$. $d^t_i$ denotes the size of data to be uploaded by AR vehicle $i \in \mathcal{N}$ at transmission iteration $t$. We denote the AR vehicle-MMBS assignment to be  $\textbf c^{t}=(c_1^t, ..., c_N^t)$, where $c_i^t=v (i \in \mathcal{N}, v \in \mathcal{M})$ denotes that AR vehicle $i$ is allocated to MMBS $v$ at iteration $t$. Considering the intra-MMBS interference, the \textit{signal to interference plus noise ratio} of AR vehicle $i$ at iteration $t$ is defined as:
\begin{align}
\Gamma_i^t(\boldsymbol{c^{t},h^{t}}) &=\frac{g^t_{i,c_i^t}h^{t}_{i}}{   \sum_{n \in\mathcal{N} \setminus \{i\}:c_n^t=c_i^t} (g^t_{n,c_i^t} h^{t}_{n}) + B\sigma^{2}}, \nonumber
\end{align}
where $h^t_{i}$ is the power of AR vehicle $i$ used for the transmission of physical world scenes to MMBS $c_i^t$ at iteration step $t$, $g^t_{c_i^t,i}$ is the channel gain between MMBS $c_i^t$ and AR vehicle $i$ at iteration step $t$. $h^t_{n}$ is the power of AR vehicle $n$ for communicating with MMBS $c_i^t$ at iteration $t$, $B$ is the bandwidth of the communicatiion, and $\sigma^2$ denotes the additive white Gaussian background noise.

Principally, $g^t_{i,c_i^t}h^{t}_{i}$ is the received signal at MMBS $c_i^t$ from AR vehicle $i$ in iteration $t$, $\sum_{n \in\mathcal{N} \setminus \{i\}:c_n^t=c_i^t} (g^t_{n,c_i^t} h^{t}_{n})$ is the intra-cell interference caused by other AR vehicle $n \neq i$ assigned to the same MMBS $c_i^t$, to AR vehicle $i$ at iteration $t$. In each iteration step $t$, the uplink data transfer rate $r^{t}_i$ from the AR vehicle $i$ to its assigned MMBS is influenced by the SINR as such:
\begin{align}
&r_{i}^{t}(\boldsymbol{c}^{t},\boldsymbol{h}^{t})=B\cdot \log_2 \left(1+\Gamma_i^t(\boldsymbol{c}^{t},\boldsymbol{h}^{t})\right),\label{eq:R2}
\end{align}
From Equation (\ref{eq:R2}), it is evident that the assignment of many AR vehicles to a single MMBS causes large intra-cell interference. A larger intra-cell interference would result in lower effective signals between AR vehicles and MMBS, and this causes the overall data transmission rate to decline. For a fixed data size to be transmitted, a higher data transfer rate results in a shorter uplink transmission delay at iteration step $t$ as shown: $\ell^{t}_i = \frac{d^{t}_i }{r^{t}_i}$, where $d^{t}_i$ is the size of the physical world scene to be uploaded from AR vehicle $i$ to a MMBS at iteration step $t$. We consider the transmitted physical world scenes to be square-frames, where data size $d^{t}_i$ and resolution $p^{t}_i$ captured by AR vehicles $i$ at iteration $t$ are related by: $d^{t}_i = \xi\cdot(p^{t}_i)^2$. $\xi$ represents the number of bits of information embedded within each pixel. Intuitively, as AR vehicle $i$ uplink latency at iteration step $t$ increases, the lower the consistency of update to the virtual world. Furthermore, a more efficient AR vehicle to MMBS allocation would increase each AR vehicles' SINR and consequently result in lower latency. Finally, the transmission of physical environment scenes of lower resolution reduces uplink transmission latency.

\begin{figure}[t]
    \centering
    \includegraphics[width=0.9\linewidth]{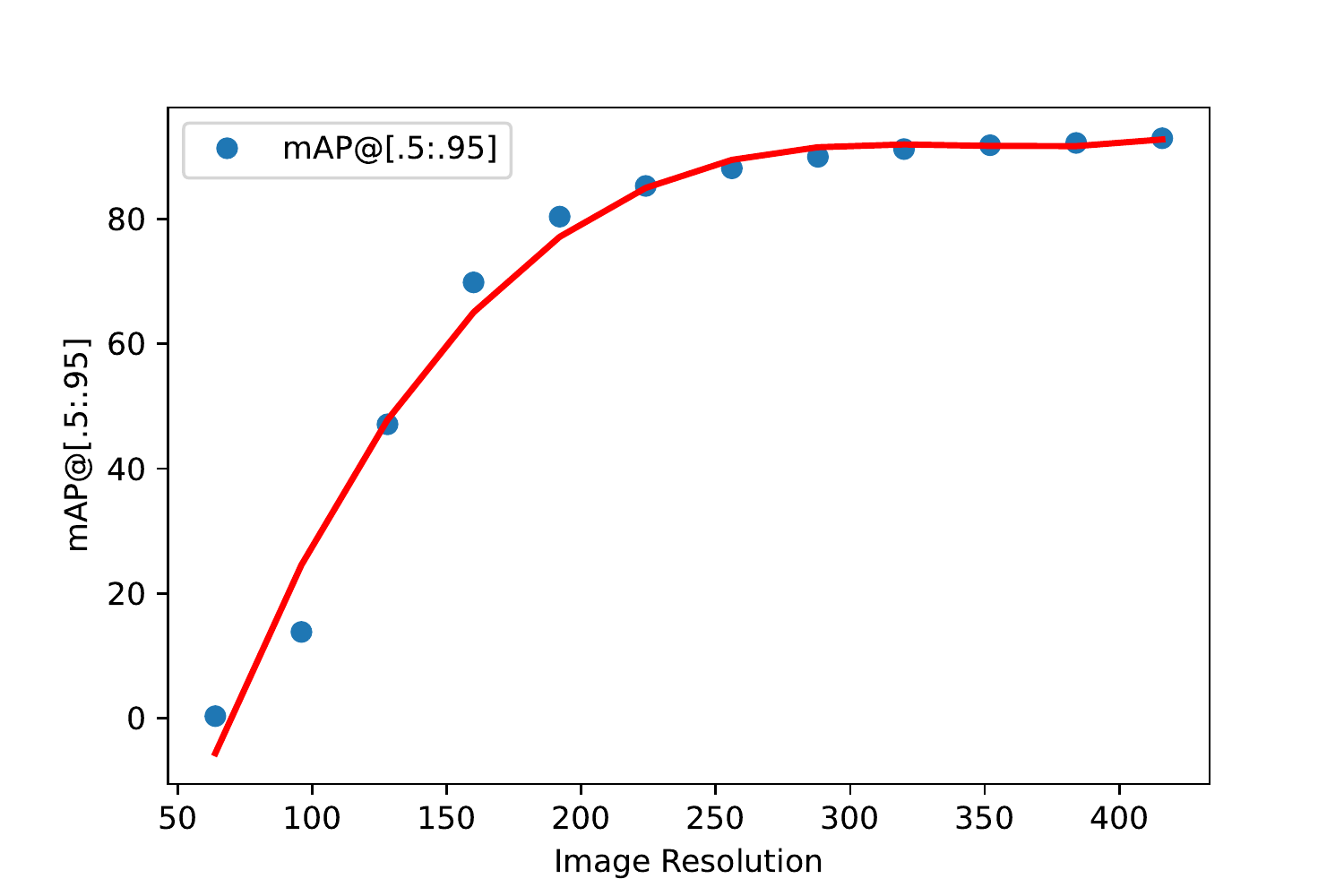}
    \caption{mAp vs Resolution (pixel per inch, ppi).}
    \label{fig:maprelationship}
    \vspace{-0.5cm}
\end{figure}

\textbf{Detection mAP-resolution model. }
As the actual implementation of continuous real-time detection of adversarial patches is infeasible for the scale of our work, we established the relationship between physical environment image resolution and adversarial patch detection mean average precision (mAP) score empirically . We collected multiple resolution-mAP pairs from the YOLOv4 prediction output and fitted a polynomial best-fit curve to the data points (as shown in Fig.~\ref{fig:maprelationship}). We note that the image resolution $p$ and mAp are related by a polynomial relationship of $\text{mAP} = 4.5\times 10^{-6}\cdot p^3 - 4.7\times10^{-3}\cdot p^2+1.6\cdot p-90$, for $p \in[64,416]$ pixel per inch (ppi).

\textbf{Idle Count. }
To ensure that we have consistent physical scene transmission for patch detection from IoVs to the MMSP, we aim to reduce the total IoVs' idle count $\sum_{t=1}^{T}\sum_{i \in \mathcal{N}}I^t_i$, which refers to minimizing the total counts in which IoVs are not uploading physical world scenes to the MMBS.

\subsection{Problem formulation} \label{problemform}
To sum up, the goal of the MMSP is to find the optimal IoV-MMBS allocation arrangement $c^{t}$ and transmitted physical environment image resolutions $p^{t}$  which minimizes the total up-link latency $\ell^{t}$ and IoV idle count $I^{t}$ while maximizing the IoV patch detection mAP $\text{mAP}(p^{t})$.  We formulated our up-link utility function as:
\begin{align}
\setlength{\belowdisplayskip}{4pt plus 1pt minus 1.0pt}
\setlength{\belowdisplayshortskip}{4pt plus 1pt minus 1.0pt}
\setlength{\abovedisplayskip}{4pt plus 1pt minus 1.0pt} \setlength{\abovedisplayshortskip}{0.0pt plus 2.0pt}
\min_{\boldsymbol{c^{t},p^{t}}} &\sum_{t=1}^{T} \sum_{i\in\mathcal{N}} q \cdot \ell^{t}_i - b\cdot \text{mAP}(p^{t}_i) + f\cdot I^{t}_{i}, \label{eq:M1}\\
s.t.~~& c_{i}^{t}\in \mathcal{M}, ~\forall i \in \mathcal{N}, \forall t \in \mathcal{T},\\
& h_{i}^t \leq h_{\max},  ~\forall i \in \mathcal{N}, \forall t \in \mathcal{T},\\
& p_{min} \leq p_i^t \leq p_{max},  ~\forall i \in \mathcal{N}, \forall t \in T
\end{align}
where $T$ is the total number of uplink transmissions of physical environment scenes from the IoVs to the MMSPs. The constraint (5) restricts each IoV to be allocated to at most one MMBS in each iteration step. Constraint (6) ensures that AR vehicle power output lies below $h_{\max}$. Constraint (7) ensures the image resolution lies between $p_{min}$ and $p_{max}$ ppi. $b,f,q$ are scaling factors which seeks to balance the order and unit difference between $\ell^t_i$ , $I^t_i$ and $\text{mAP}(p^{t}_i)$ for joint-variable optimization.

\section{Reinforcement learning settings}
\label{RL}
For our work, we assign two reinforcement learning agents, $Agent1$ and $Agent2$, with $Agent1$ performing the discrete action of IoV-MMBS allocation, and $Agent2$ performing the continuous action IoV uplink image resolution selection. Both agents are incorporated within the MMSP and represent the MMSP's interests. The rationale for adopting two agents is that we are optimizing two variables in which one has continuous and the other has discrete action spaces.

\textbf{State. }For both agents' observation state $s^t$, we have chosen to include 1) \textbf{channel gain between each IoV and all MMSPs}: $g^t_{v,i}$, 2) \textbf{image data size to be transmitted by each IoV} at each transmission iteration $t$: $d^{t}_{i}$, as these two variables influences data up-link transmission rate and latency.

\textbf{Action. }For $Agent1$, the agent's action is to decide the MMBS to IoV allocation, in which the action space for each IoV $i$ can be written as such: $a^{alloc,t}~=~\boldsymbol{c}^{t}~=~(c^{t}_{1},...,c^{t}_{N}),\\~(t\in T)$. The number of the discrete actions is $N^{M+1}$, where $N$ denotes the number of IoVs and $M$ is the total number of MMBSs. This signifies that an IoV may or may not be allocated to a MMBS.

For $Agent2$, the action space is continuous and the action dimension is $N$, in which there is one image resolution value selected for each IoV to transmit the physical environment scenes to its assigned MMBS. The uplink action space for all the IoVs at each transmission iteration is written as such: $a^{resol,t}=\boldsymbol{p}^{t}=\{p^t_{1}, ... ,p^t_{N}\},~(t\in T).$, where $p^t_{i=N}$ is the uplink image resolution selected at iteration $t$ for IoV $N$.


\textbf{Reward. }
Although we have a single objective function, in practice, we break down the overarching objectives into smaller rewards to be assigned to each of our agents. We assign only components of the objective function which is influenced by an agent's decision to that agent.

For the $Agent1$ the reward is given at transmission iteration $t$ as such:
\begin{align}
    \mathcal{R}^{alloc,t} = - \frac{ \sum_{i \in \mathcal{N}}\left(q\cdot\ell^{t}_i + f\cdot I^{t}_i\right)}{N}
    \label{eq:reward_down}
\end{align}
while for $Agent2$, the reward given to the agent at transmission iteration $t$ is given as such:
\begin{align}
    \mathcal{R}^{resol,t} = -\frac{ \sum_{i \in \mathcal{N}}q \cdot \ell^{t}_i - b\cdot \text{mAP}(p^{t}_i)}{N}
    \label{eq:reward_up}
\end{align}
We divide the reward functions by $N$ IoVs to find an average reward, as the average reward received per IoV is more intuitive than the reward sum.

\subsection{Heterogeneous Actions PPO}
 \label{algorithms}
 Inspired by the well-known Centralized Training Decentralized Execution (CTDE) framework~\cite{lowe2017multi}, we developed the Heterogeneous Actions Proximal Policy Optimization (HAPPO) algorithm for our dual-agent RL model. This model features both discrete and continuous action spaces and utilizes PPO as the backbone, as PPO is considered a state-of-the-art algorithm with performance stability. We do not directly use traditional CTDE algorithms like Multi-Agent PPO (MAPPO) because the actions in our scenario contain both discrete and continuous actions, and it is not feasible to directly concatenate them to form a unified action. This is because the discrete action space PPO and the continuous action space PPO use different networks and distributions for sampling actions. 
 
Similar to PPO~\cite{PPO}, HAPPO uses separate policies $\pi_\theta$ and $\pi_{\theta^{'}}$ for sampling trajectories (during training) and evaluation, respectively. Here, $\pi_{\theta_1}$ and $\pi_{\theta_2}$ are two separate distributions instead of a shared distribution in policy optimization. KL divergence constraints are applied to both Actors' policies to prevent major policy changes in each update. As the Actor-network is based on policy gradient~\cite{sutton1999policy}, according to PPO~\cite{PPO}, we formulate the update function of Actors as:
\begin{align}
    \mathbb{E}_{(s^t,a^{alloc,t})\sim\pi_{\theta_1^{'}}}[f^{alloc,t}(\theta_1)(A^{alloc,t}+A^{resol,t})] \\
    \mathbb{E}_{(s^t,a^{resol,t})\sim\pi_{\theta_2^{'}}}[f^{resol,t}(\theta_2)(A^{alloc,t}+A^{resol,t})]
\end{align}
where
\begin{align}
    &f^{alloc,t}(\theta_1)=\text{min}\{\mathcal{R}^{alloc,t}(\theta_1), \text{clip}(\mathcal{R}^{alloc,t}(\theta_1), 1-\epsilon, 1+\epsilon)\}
\end{align}
\begin{align}
    &\text{and}~~~\mathcal{R}^{alloc,t}(\theta_1)=\frac{\pi_{\theta_1}(a^{alloc,t}|s^t)}{\pi_{{\theta'_1}}(a^{alloc,t}|s^t)}. 
\end{align}
$f^{resol,t}(\theta_2)$ and $\mathcal{R}^{resol,t}(\theta_2)$ is also defined in the same manner as equation (13), (14), respectively, with $resol$ replacing $alloc$ in the superscripts. $\epsilon$ refers to the policy clipping parameter.
 
Here, $A^{alloc,t}$ and $A^{resol,t}$ are the advantages of actions selected by $Agent1$ and $Agent2$, respectively. The advantages are computed by the truncated version of TD($\lambda$)~\cite{schulman2015high}. 
\begin{align}
    &A^{alloc,t} = \delta^{alloc,t} + ...+(\gamma\lambda)^{\bar{T}-1}\delta^{alloc,t+\bar{T}-1}, \\
    &\text{where}~~~\delta^{alloc,t}=\mathcal{R}^{alloc,t}+\gamma V_{\phi'}(s^{t+1})-V_{\phi'}(s^t).
\end{align}
$A^{resol,t}$ and $\delta^{resol,t}$ is defined in the same manner as equation (15), (16), respectively, with $resol$ replacing $alloc$ in the superscripts. $\bar{T}$ is the trajectory segment, $\lambda$ is the trace decay parameter and $\gamma$ is the discount rate.

In terms of the value network (Critic), HAPPO uses identical Critics as per other Actor-Critic algorithms; and the loss function can be formulated as:
\begin{align}
    L(\phi) = [V_\phi(s^t)-((A^{alloc,t}+A^{resol,t})+\gamma V_{\phi'}(s^{t+1}))]^2 \label{eq:criticloss}
\end{align}

where $V(s)$ is the widely used state-value function~\cite{RLintro}, which is estimated by a learned critic network with parameter $\phi$. We update $\phi$ by minimizing the $L(\phi)$, and the parameter $\phi'$ of target state-value function periodically with $\phi$.

\begin{figure}[t]
    \centering
    \includegraphics[width=1\linewidth]{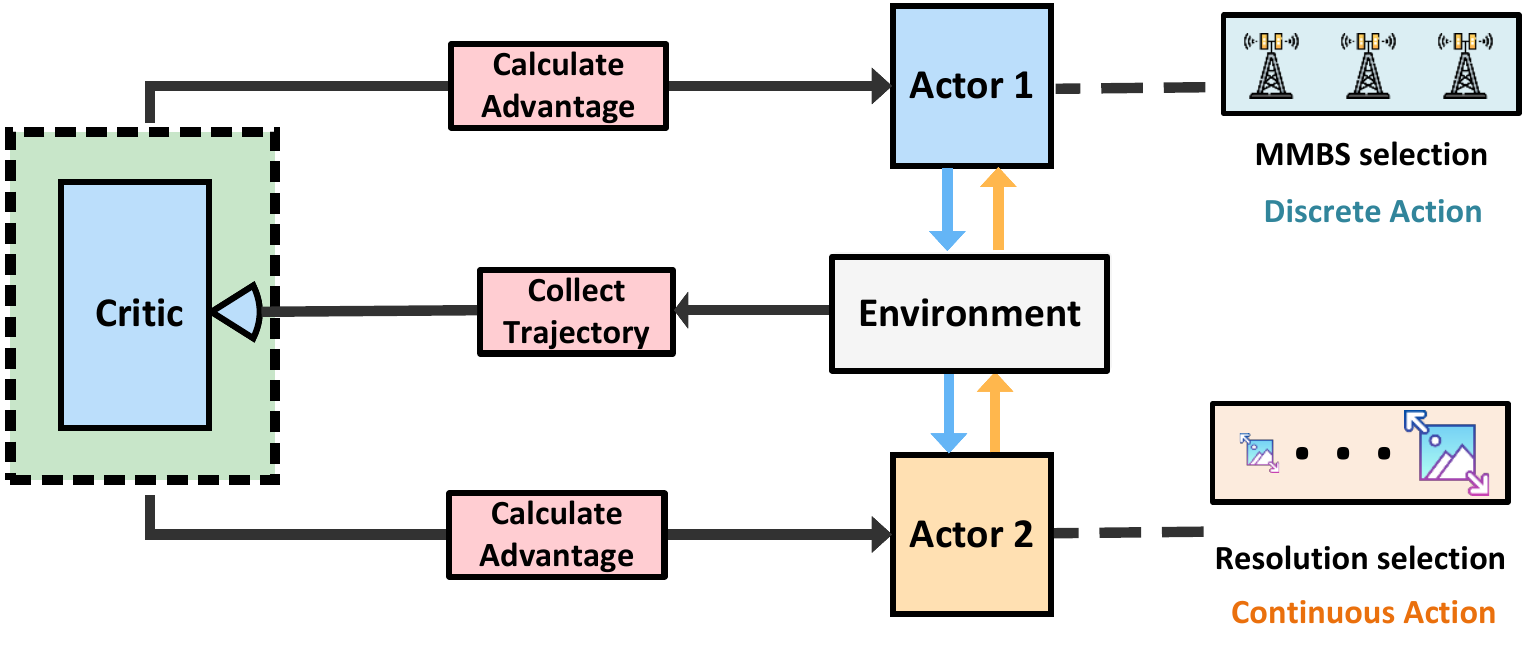}
    \caption{Heterogeneous Action PPO (HAPPO) structure.}
    \label{fig:algorithm}
    \vspace{-0.5cm}
\end{figure}

\begin{figure}[!t] 
        \renewcommand{\algorithmicrequire}{\textbf{Initiate:}}
        \renewcommand{\algorithmicensure}{\textbf{Output:}}
        \begin{algorithm}[H]
            \caption{\label{alg:PPO}Heterogenous Action PPO}
            \begin{algorithmic}[1]
                \REQUIRE critic parameter $\phi$ and target network $\phi^{'}$, $Agent1$ actor parameter $\theta_{1}$, $Agent2$ actor parameter $\theta_{2}$, initialize state $s^{t}=s^{1}$
                \FOR{iteration = $1,2,...$}
                    \STATE $Agent1$ and $Agent2$ execute action according to $\pi_{\theta^{'}_{1}}(a^{alloc,t}|s^{t})$ and $\pi_{\theta^{'}_{2}}(a^{resol,t}|s^{t})$, respectively
                    \STATE Get $\mathcal{R}^{alloc,t}$ and $\mathcal{R}^{resol,t}$ and next state $s^{t+1}$
                        
                    \STATE sample trajectories: \\$\tau$=$\{s^{t},a^{alloc,t},a^{resol,t},s^{t+1},\mathcal{R}^{alloc,t},\mathcal{R}^{resol,t}\}$ iteratively
                    \STATE Compute advantages $\{A^{alloc,t},A^{resol,t}\}$
                    \STATE Compute target values \{$V^{alloc,t}_{targ},V^{resol,t}_{targ}$\}
                    \FOR{$k$ = $1,2,...,K$}
                        \STATE Shuffle the data's order, set batch size $bs$
                        \FOR{$j$=$0,1,...,\frac{T}{bs}-1$}
                            \STATE Compute gradient for downlink and uplink actors:
                            $\triangledown \theta_{1}, \triangledown \theta_{2}$
                            \STATE Apply gradient ascent on $\theta_{1}$ using $\triangledown \theta_{1}$
                            \STATE Apply gradient ascent on $\theta_{2}$ using $\triangledown \theta_{2}$ 
                            \STATE Update critic with loss using eq.~(\ref{eq:criticloss})
                        \ENDFOR
                        \STATE Assign target network parameters $\phi^{'} \leftarrow \phi$ after $C$ iterations
                    \ENDFOR
                    
                \ENDFOR
            \end{algorithmic}
        \end{algorithm}
\vspace{-0.8cm}
\end{figure}
\section{Experiment}
\label{experiment}
In this section, we will describe our experimental configurations and provide in-depth analyses on the results.
\subsection{Configuration} \label{Configuration}
We use five congestion settings: 3 MMBS and with 3 to 7 IoVs (denoted as $"3x"$ for 3 MMBS and $x$ number of IoVs) to test our proposed HAPPO orchestrator. We compared (i) our proposed HAPPO against baseline models (ii) Independent Dual Agent PPO-PPO, (iii) Heterogeneous A2C (HAA2C) which utilizes similar structure to HAPPO, and (iv) a random IoV-MMBS allocation and image resolution selection agent. The bandwidth and noise are simulated to be $B=10$ MHz and $\sigma^2=-100$ dBm. We initialize and constrain IoV power output, image resolution, and IoV locations for different IoVs to $(1.5,2.0)$ Watt, $(64,416)$ ppi, $x,y\in(0,1000)$ m, respectively. $x$ and $y$ represents the relative longitudinal and latitudinal directions in our 1000m by 1000m map, and IoVs randomly move a maximum of 100m in $x$ and $y$ directions in each transmission iteration. We set $b$, $q$ and $f$ to be 50, 60 and 75, respectively, and these numbers are empirically derived. We adopt the ADAM optimizer\cite{adam} for all our implemented algorithms. To better observe the final performance, we use 280,000 steps for training. We conducted the training and simultaneous evaluation of the models for each of the configurations at different seed settings: seed 0 to seed 9.



\begin{figure}[t]
\centering
\subfigtopskip=2pt
\subfigbottomskip=2pt

\subfigure[Agent1 reward in 34 scenario.]{
\begin{minipage}[t]{0.8\linewidth}
\centering
\includegraphics[width=1\linewidth]{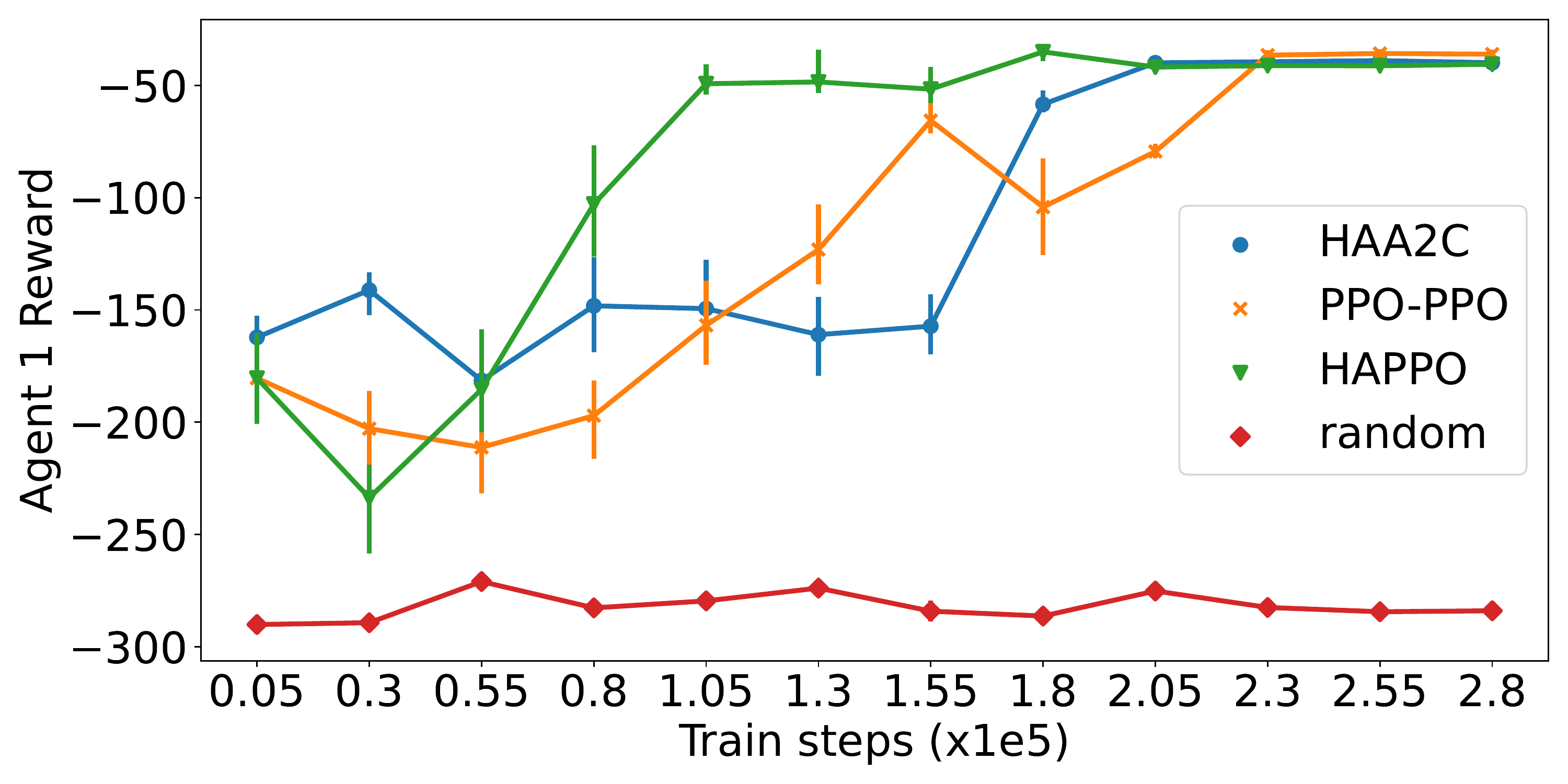}
\label{fig:34_agent1}
\vspace{-10mm}
\end{minipage}
}%

\subfigure[Agent2 reward in 34 scenario.]{
\begin{minipage}[t]{0.8\linewidth}
\centering
\includegraphics[width=1\linewidth]{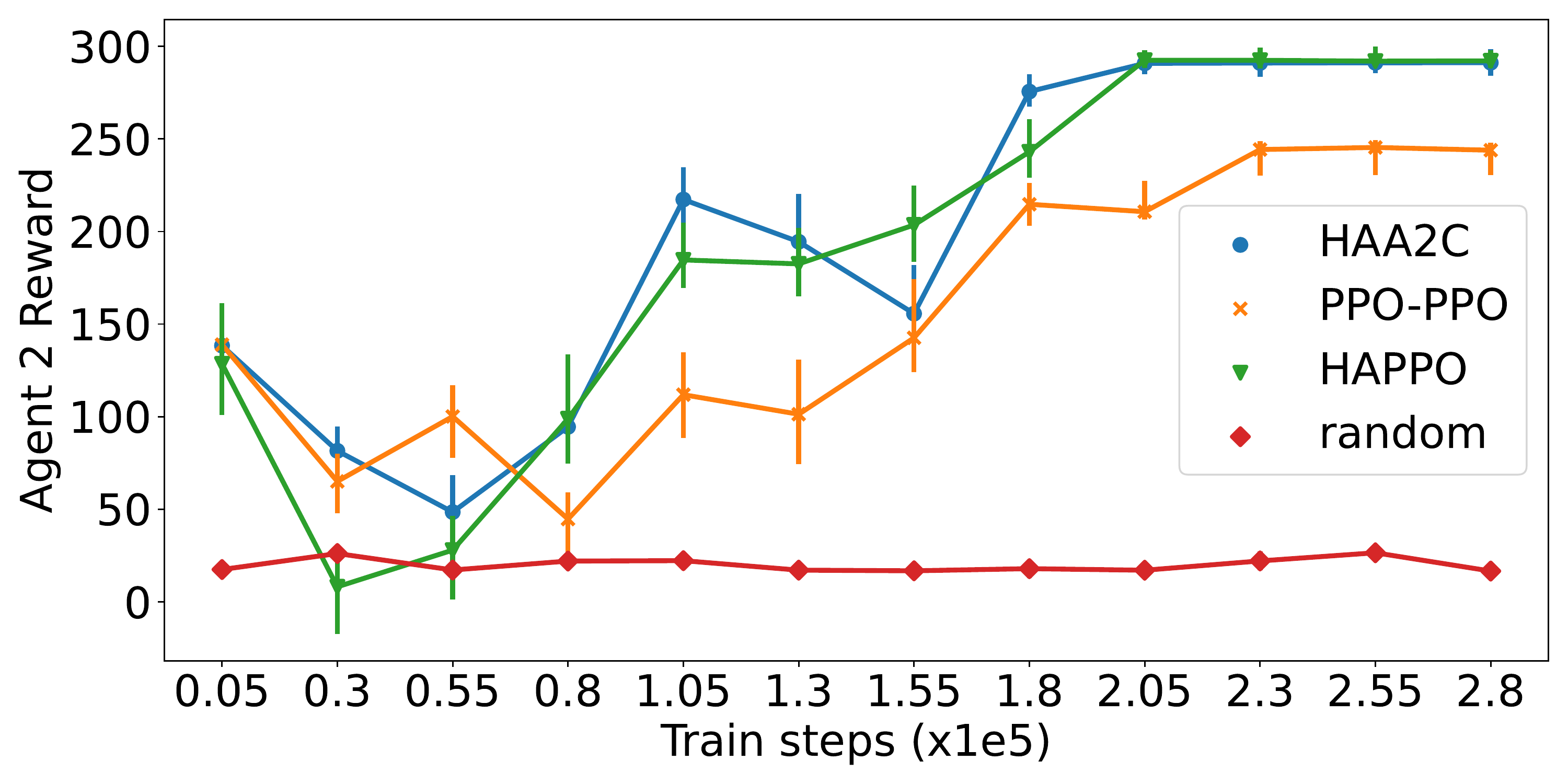}
\label{fig:34_agent2}
\vspace{-10mm}
\end{minipage}
}%

\subfigure[Agent1 reward in 37 scenario.]{
\begin{minipage}[t]{0.8\linewidth}
\centering
\includegraphics[width=1\linewidth]{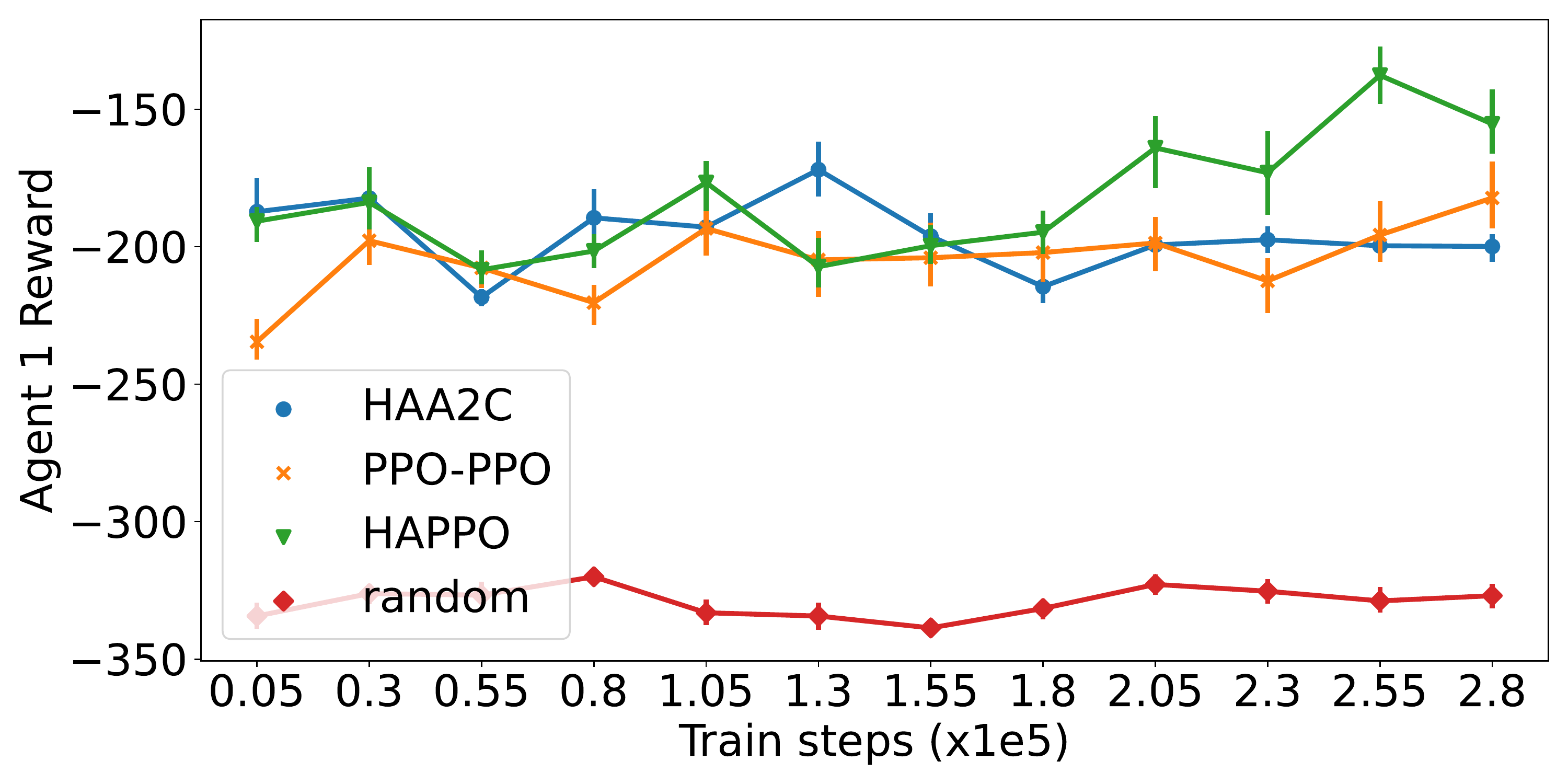}
\label{fig:37_agent1}
\vspace{-10mm}
\end{minipage}%
}%

\subfigure[Agent2 reward in 37 scenario.]{
\begin{minipage}[t]{0.8\linewidth}
\centering
\includegraphics[width=1\linewidth]{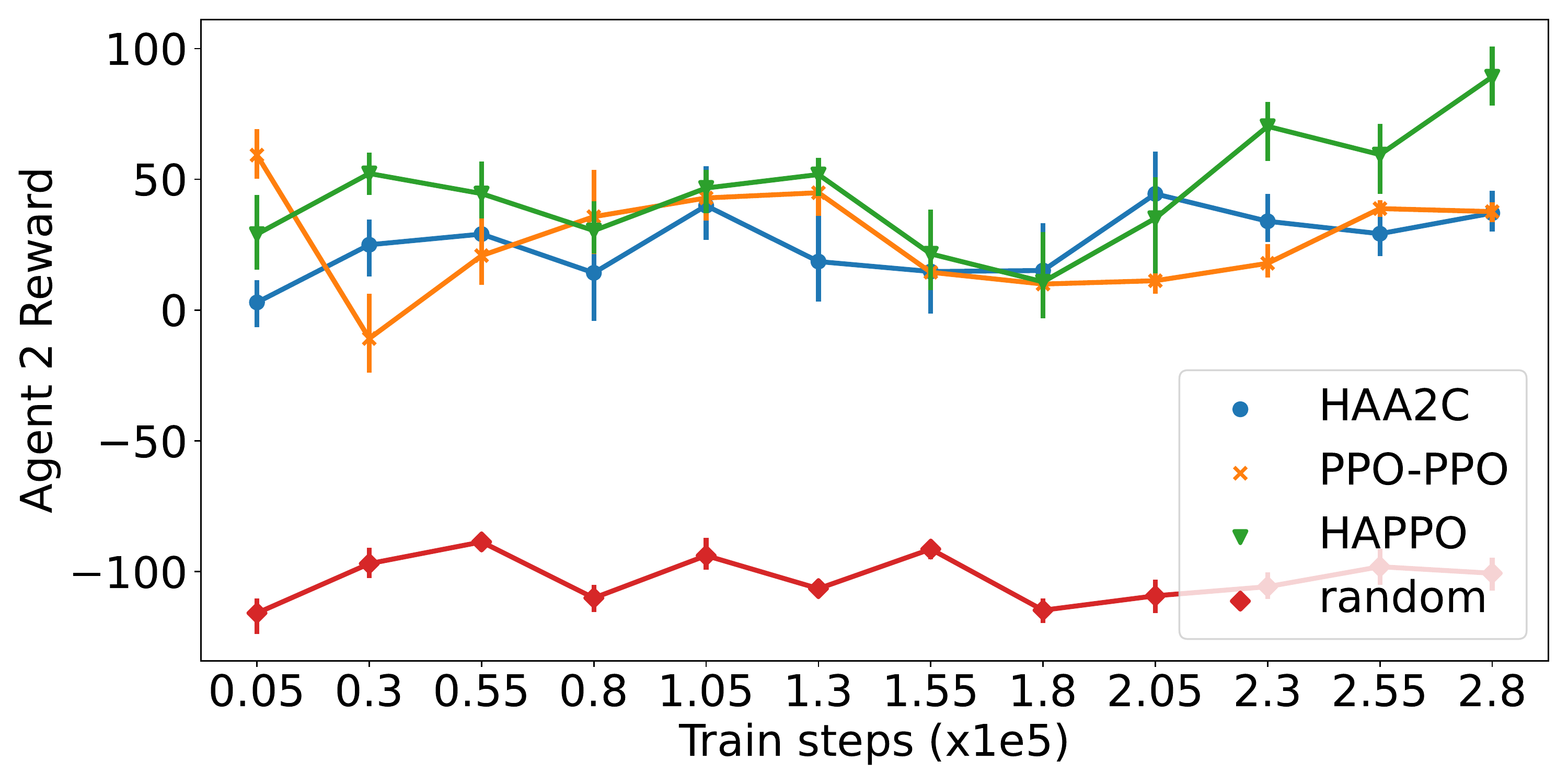}
\label{fig:37_agent2}
\vspace{-10mm}
\end{minipage}%
}%

\caption{Reward during training in scenario 34 and 37.}
\vspace{-0.5cm}
\end{figure}

\begin{figure}[t]
    \centering
    \includegraphics[width=0.9\linewidth]{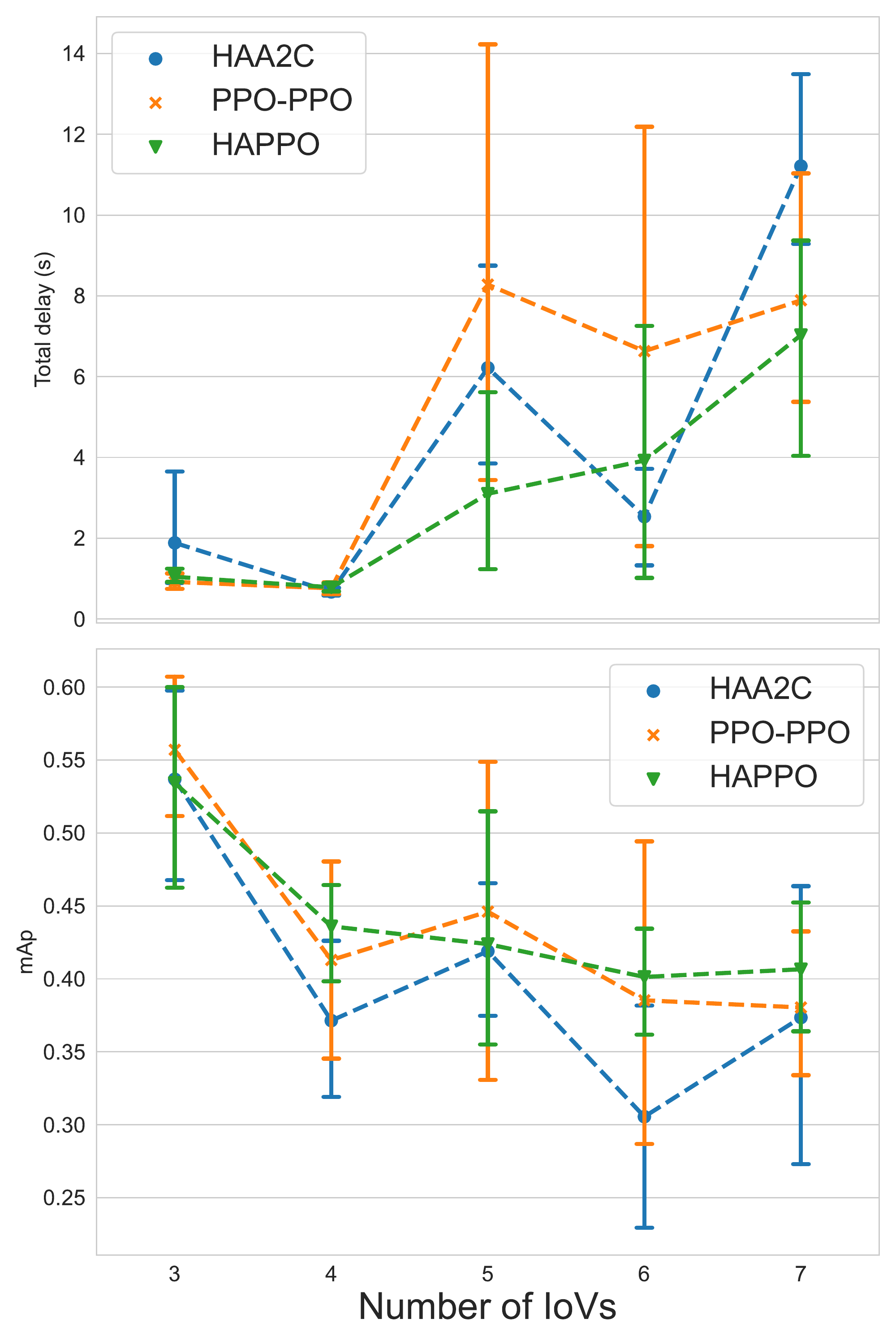}
    \caption{Metrics with different User numbers.}
    \label{fig:metrics}
    \vspace{-0.5cm}
\end{figure}

\begin{table}[t]
\centering
\caption{Overall rewards}
\label{table:parameter}
\vspace{-0.2mm}
\scalebox{1}{
\begin{tabular}{cccc}
\hline
Number of IoV & \makecell{HAPPO} & \makecell{HAA2C} & \makecell{Independent agents\\ (PPO-PPO)} \\ \hline
\multicolumn{4}{c}{Agent1 Reward} \\ \hline
$3$ & $-47.53$ & $-$\textbf{43.82} & $-48.94$\\
$4$ & $-42.27$ & $-41.10$ & $-$\textbf{40.36}\\
$5$ & $-$\textbf{132.23} & $-160.45$ & $-154.83$ \\
$6$ & $-$\textbf{148.34} & $-173.42$ & $-165.78$ \\
$7$ & $-$\textbf{154.47} & $-197.47$ & $-184.23$ \\ \hline
\multicolumn{4}{c}{Agent2 Reward} \\ \hline
$3$ & \textbf{363.68} & $344.59$ & $360.36$ \\
$4$ & \textbf{287.43} & $286.72$ & $238.58$\\
$5$ & \textbf{234.72} & $152.53$ & $163.49$\\
$6$ & \textbf{169.34} & $134.29$ & $145.6$ \\
$7$ & \textbf{88.74} & $38.10$ & $39.34$\\ \hline
\end{tabular}
}
\label{table:results}
\vspace{-0.5cm}
\end{table}

\subsection{Result analyses}
We present the final obtained rewards for both $Agent1$ and $Agent2$ in Table~\ref{table:results}. In the simpler "33" and "34" settings, most of the RL algorithm pairs we adopted performed fairly well and achieved convergence, with the exception of independent PPO-PPO in the "34" setting (reflected in poorer rewards obtained shown in Fig.~\ref{fig:34_agent2}). Nevertheless, we observed a notably quicker training convergence by the proposed HAPPO algorithm (shown in Fig.~\ref{fig:34_agent1}). In the more complex scenarios such as "35", "36", and "37", we found that our adopted HAPPO achieved significantly better rewards than the other baseline RL algorithms (shown in Table~\ref{table:results} and Fig.~\ref{fig:37_agent1} and Fig.~\ref{fig:37_agent2}).

The total up-link transmission (of a batch of 1000 scenes) increases substantially as the number of IoVs increase, while the mAP score achieved decreased as the number of IoVs increased (shown in Fig.~\ref{fig:metrics}). This is not unexpected as the more complex scenarios involve more IoVs sharing computing resources with an unchanging number of MBBS. HAPPO showcased its superiority over the other algorithm by obtaining the lowest average total transmission delay and considerably good mAP score, across the different congestion settings. Furthermore, HAPPO exhibits a much narrower range of total uplink transmission time and mAP score (as shown by the error bars in Fig.~\ref{fig:metrics}) when compared to other algorithm, indicating greater stability.

Despite the disparities in performance between the different algorithms, all algorithm performed better than an agent which allocates IoV-MBBS and selects uploaded image resolution randomly (shown in Fig.~\ref{fig:34_agent1},~\ref{fig:34_agent2},~\ref{fig:37_agent1},~\ref{fig:37_agent2}). This further substantiates that our proposed orchestrator improves the uplink communication in terms of maximizing accuracy, minimizing transmission delay and IoV idle counts.

\section{Conclusion}
In our work, we have proposed a real-time adversarial patch detector, enabled by mobile edge computing, in the defence of digital twinning to the metaverse. We formulated a realistic joint variable optimization problem where the MMSPs' objective is to maximize adversarial patch detection mAP, while minimizing the uplink transmission latency and IoV idle counts, through optimizing the MMBS allocation and IoV uplink image resolution. We proposed a Heterogenous Action PPO (HAPPO) (discrete-continuous) to tackle our proposed problem. We have demonstrated that our proposed HAPPO model outperforms baseline models and achieved superior performance based on key metrics.

\section*{Acknowledgement}

This research is partly supported by the Singapore Ministry of Education Academic Research Fund under Grant Tier 1 RG90/22, RG97/20, Grant Tier 1 RG24/20 and Grant Tier 2 MOE2019-T2-1-176; and partly by the NTU-Wallenberg AI, Autonomous Systems and Software Program (WASP) Joint Project.

{\small
\bibliographystyle{IEEEtran}

}

\end{document}